\documentclass[9pt,twocolumn,twoside]{pnas-new}
\templatetype{pnasresearcharticle}

\begin{document}

% \doublespacing % For draft

\title{Image-Space Gridding for Nonrigid Motion-Corrected MR Image Reconstruction}

% Use letters for affiliations, numbers to show equal authorship (if applicable) and to indicate the corresponding author
\author[a,1]{\orcid{0009-0000-7031-1347}{Kwang~Eun~Jang}}
\author[a,b]{\orcid{0000-0003-0063-564X}{Mario~O.~Malav\'{e}}}
\author[a]{\orcid{0009-0005-5963-765X}{Dwight~G.~Nishimura}}

\affil[a]{Magnetic Resonance Systems Research Lab (MRSRL), Department of Electrical Engineering, Stanford University}
\affil[b]{Currently affiliated with Apple Inc.}

% Please give the surname of the lead author for the running footer
\leadauthor{Jang}

% Please add a significance statement to explain the relevance of your work
% Authors must submit a 120-word maximum statement about the significance of their research paper written at a level understandable to an undergraduate educated scientist outside their field of speciality. The primary goal of the significance statement is to explain the relevance of the work in broad context to a broad readership. The significance statement appears in the paper itself and is required for all research papers.
\significancestatement{
    Motion remains a major challenge in magnetic resonance imaging, particularly in free-breathing cardiac imaging, where data are acquired over multiple heartbeats. Here, we derive image-space gridding by adapting the nonuniform fast Fourier transform (NUFFT) to represent and compute nonrigid motion. Additionally, we present a method that employs both low-resolution 3D image-based navigators and high-resolution 3D self-navigating image-based navigators to estimate translational and nonrigid motion, respectively. Using image-space gridding and the proposed motion estimation scheme, we concisely reformulate the nonrigid motion correction problem as a standard regularized inverse problem. This work could be particularly beneficial for patients unable to hold their breath during cardiac MR scans.
}

% Please include corresponding author, author contribution and author declaration information
\authorcontributions{K.E.J. developed the theoretical framework, implemented the algorithm, analyzed the data, and wrote the manuscript. M.O.M. conducted in-vivo experiments. D.G.N. supervised the project and revised the manuscript. All authors discussed the results and contributed to the final version of the manuscript.}

% \authordeclaration{This study was funded by NIH Grant HL127039.}
% \equalauthors{\textsuperscript{1}A.O.(Author One) contributed equally to this work with A.T. (Author Two) (remove if not applicable).}
\correspondingauthor{\textsuperscript{1}To whom correspondence should be addressed. E-mail: kejang@stanford.edu}

% At least three keywords are required at submission. Please provide three to five keywords, separated by the pipe symbol.

\begin{abstract}
    Motion remains a major challenge in magnetic resonance (MR) imaging, particularly in free-breathing cardiac MR imaging, where data are acquired over multiple heartbeats at varying respiratory phases. We adopt a model-based approach for nonrigid motion correction, addressing two challenges: (a) motion representation and (b) motion estimation. For motion representation, we derive image-space gridding by adapting the nonuniform fast Fourier transform (NUFFT) to represent and compute nonrigid motion, which provides an exact forward-adjoint pair of linear operators. We then introduce nonrigid SENSE operators that incorporate nonrigid motion into the multi-coil MR acquisition model. For motion estimation, we employ both low-resolution 3D image-based navigators (iNAVs) and high-resolution 3D self-navigating image-based navigators (self-iNAVs). During each heartbeat, data are acquired along two types of non-Cartesian trajectories: a subset of a high-resolution trajectory that sparsely covers 3D k-space, followed by a full low-resolution trajectory. We reconstruct 3D iNAVs for each heartbeat using the full low-resolution data, which are then used to estimate bulk motion and identify the respiratory phase of each heartbeat. By combining data from multiple heartbeats within the same respiratory phase, we reconstruct high-resolution 3D self-iNAVs, allowing estimation of nonrigid respiratory motion. For each respiratory phase, we construct the nonrigid SENSE operator, reformulating the nonrigid motion-corrected reconstruction as a standard regularized inverse problem. In a preliminary study, the proposed method enhanced sharpness of the coronary arteries and improved image quality in non-cardiac regions, outperforming translational motion-corrected reconstruction.
\end{abstract}

\dates{This manuscript was compiled on \today}
% \doi{\url{www.pnas.org/cgi/doi/10.1073/pnas.XXXXXXXXXX}}

\maketitle
\thispagestyle{firststyle}
\ifthenelse{\boolean{shortarticle}}{\ifthenelse{\boolean{singlecolumn}}{\abscontentformatted}{\abscontent}}{}

\firstpage[5]{4}
% \firstpage[5]{2}    % For draft

% Use \firstpage to indicate which paragraph and line will start the second page and subsequent formatting. In this example, there are a total of 11 paragraphs on the first page, counting the first level heading as a paragraph. The value {12} represents the number of the paragraph starting the second page. If a paragraph runs over onto the second page, include a bracket with the paragraph line number starting the second page, followed by the paragraph number in curly brackets, e.g. "\firstpage[4]{11}".

% If your first paragraph (i.e. with the \dropcap) contains a list environment (quote, quotation, theorem, definition, enumerate, itemize...), the line after the list may have some extra indentation. If this is the case, add \parshape=0 to the end of the list environment. 
% Note: please start your introduction without including the word ``Introduction'' as a section heading (except for math articles in the Physical Sciences section); this heading is implied in the first paragraphs.

\dropcap{M}agnetic resonance (MR) imaging scans generally consist of multiple repetitions of pulse sequences, with each repetition typically lasting from a few milliseconds to several hundred milliseconds. Since movement during this scan period introduces inconsistencies in k-space, often manifesting as ghosting or blurring in image-space, motion remains a major challenge in MR imaging.

We adopt a model-based approach \cite{odille2008generalized} for nonrigid motion correction, formulating an optimization problem that accounts for subject motion and solving it using an iterative solver. This approach involves two primary challenges: (a) motion representation and (b) motion estimation. Motion representation in the model should meet three key requirements. First, it should describe a wide range of motion. Second, its adjoint operation should also be available. Third, it should be computable within a reasonable amount of time, as both the forward and adjoint operations are repeatedly executed within the iterative solver. Motion estimation should have minimal impact on data acquisition while providing sufficient accuracy.

For motion representation, we derive image-space gridding by adapting the nonuniform fast Fourier transform (NUFFT) \cite{dutt_fast_1993, fessler_nonuniform_2003, beatty_rapid_2005, fessler_nufft-based_2007} to represent and compute nonrigid motion. Since the proposed linear operators form an exact forward-adjoint pair that can be efficiently implemented on graphics processing units (GPUs) to accelerate computations \cite{gregerson2008implementing}, they are well-suited for iterative reconstructions. Using image-space gridding, we introduce a new linear operator, termed the nonrigid SENSE operator, by preceding the conventional SENSE operator \cite{pruessmann_sense_1999} with the forward image-space gridding operator. The nonrigid SENSE operator incorporates the nonrigid transformation of the object into the multi-coil MR acquisition model. By stacking multiple nonrigid SENSE operators corresponding to different transformations of the object, we formulate a linear equation that models the state transitions of the object during the prolonged MR scan.

For motion estimation, we present a method that employs both low-resolution 3D image-based navigators (iNAVs) \cite{wu2013free, addy20173d} and high-resolution 3D self-navigating image-based navigators (self-iNAVs) \cite{pang2014whole}, with a primary focus on addressing respiratory motion in free-breathing coronary MR angiography \cite{feng_xd-grasp_2016, wu2013free, bhat20113d, addy20173d, spuentrup_free-breathing_2004}. During each heartbeat, data are acquired along two types of non-Cartesian trajectories: a subset of a high-resolution trajectory that sparsely covers 3D k-space, followed by a full low-resolution trajectory. We reconstruct 3D iNAVs for each heartbeat using data near the k-space origin, which are then used to estimate bulk motion and identify the respiratory phase of each heartbeat. Since the high-resolution data associated with each heartbeat are distributed across 3D k-space, binning datasets within the same respiratory phase enables the reconstruction of a set of high-resolution 3D self-iNAVs across the respiratory cycle. This approach allows capturing nonrigid respiratory motion using a relatively short low-resolution iNAV acquisition during each heartbeat.

The process of nonrigid motion-corrected reconstruction is summarized as follows: A nonrigid image registration method is applied across the 3D self-iNAV to obtain displacement fields \cite{rueckert_nonrigid_1999, modersitzki2003numerical} with respect to a reference respiratory phase. Using the estimated displacement fields, the proposed nonrigid SENSE operators are constructed for each respiratory phase, incorporating nonrigid motion across multiple heartbeats into the MR acquisition model. In this way, we reformulate the nonrigid motion-corrected reconstruction as a standard regularized inverse problem. In a preliminary study, we show that the proposed method enhances voxel depiction and improves image quality in non-cardiac regions, outperforming translational motion-corrected reconstruction.

\section*{Notations}\label{sec:notations}

We use $[\cdot]$ and $(\cdot)$ to describe functions or data on Cartesian and non-Cartesian grids, respectively. For example, $M(k_x)$ represents k-space data along a non-Cartesian trajectory, while $m[x]$ denotes the image to be reconstructed on a Cartesian grid. We utilize $\{\cdot\}$ to denote the application of operators. Our symbols for gridding subtly adapt notations in Beatty et al. \cite{beatty_rapid_2005}. Table \ref{tab:notations} summarizes the notations used in this paper.

\begin{table}[t!]
    \centering
    \caption{Notations}\label{tab:notations}
    \begin{tabular}{ll}
    \toprule
    \textbf{Symbol} & \textbf{Definition} \\
    \midrule
    $x$      & Image location \\
    $k_x$    & K-space location \\
    $N$      & Number of pixels \\
    $G$      & Number of sample locations on an oversampled grid\\
    $M(k_x)$ & Sampled, density corrected k-space data \\
    $\mathrm{III}(x)$ & Comb function, $\mathrm{III}(x) = \sum_i \delta(x - i)$ \\
    $c(x)$   & Interpolation kernel function in image-space \\
    $C(k_x)$ & Interpolation kernel function in k-space \\
    $m[x]$   & Image to be reconstructed at the reference phase \\
    $m_j[x]$ & Warped image at the $j$-th phase \\
    $\mathcal{T}_j\{\cdot\}$    & Nonrigid transform that warps $m[x]$ to $m_j[x]$ \\ 
    $\mathcal{T}_j^H\{\cdot\}$  & Adjoint of $\mathcal{T}_j\{\cdot\}$ \\ 
    $\mathcal{F}\{\cdot\}$    & Fourier transform \\
    $\mathcal{F}^H\{\cdot\}$  & Inverse Fourier transform \\
    $\mathcal{Z}\{\cdot\}$    & Zero-padding \\
    $\mathcal{Z}^H\{\cdot\}$  & Undo zero-padding (cropping) \\
    $\mathcal{S}\{\cdot\}$    & Multiplying a coil sensitivity map\\
    $\mathcal{P}^{(k)}\{\cdot\}$  & Projection Core in k-space\\
    $\mathcal{P}^{(i)}\{\cdot\}$  & Projection Core in image-space\\
    $\mathcal{B}^{(k)}\{\cdot\}$  & Backprojection Core in k-space\\
    $\mathcal{B}^{(i)}\{\cdot\}$  & Backprojection Core in image-space\\
    iNAV & image-based navigator\\
    self-iNAVs & self-navigating image-based navigator\\
    \bottomrule
    \end{tabular}
\end{table}

\section*{Image-based Gridding}

We would like to clarify the term ``gridding'' as used in this paper. In the MR imaging context, ``gridding'' denotes a specific type of NUFFT; it transforms nonuniform k-space samples to an image on a Cartesian grid (Type-1 NUFFT \cite{dutt_fast_1993, fessler_nonuniform_2003, fessler_nufft-based_2007}). To avoid confusion, we label this Type-1 NUFFT as ``k-space gridding.'' Similarly, the Type-2 NUFFT is termed ``inverse k-space gridding,'' which generates k-space data on a non-Cartesian grid from an image on a Cartesian grid.

\subsection*{k-space Gridding}\label{subsec:grid-theory-k_space}

\begin{figure}[!htb]
    \centering
    \begin{minipage}{0.22\textwidth}
        \centering
        \includegraphics[width=\linewidth]{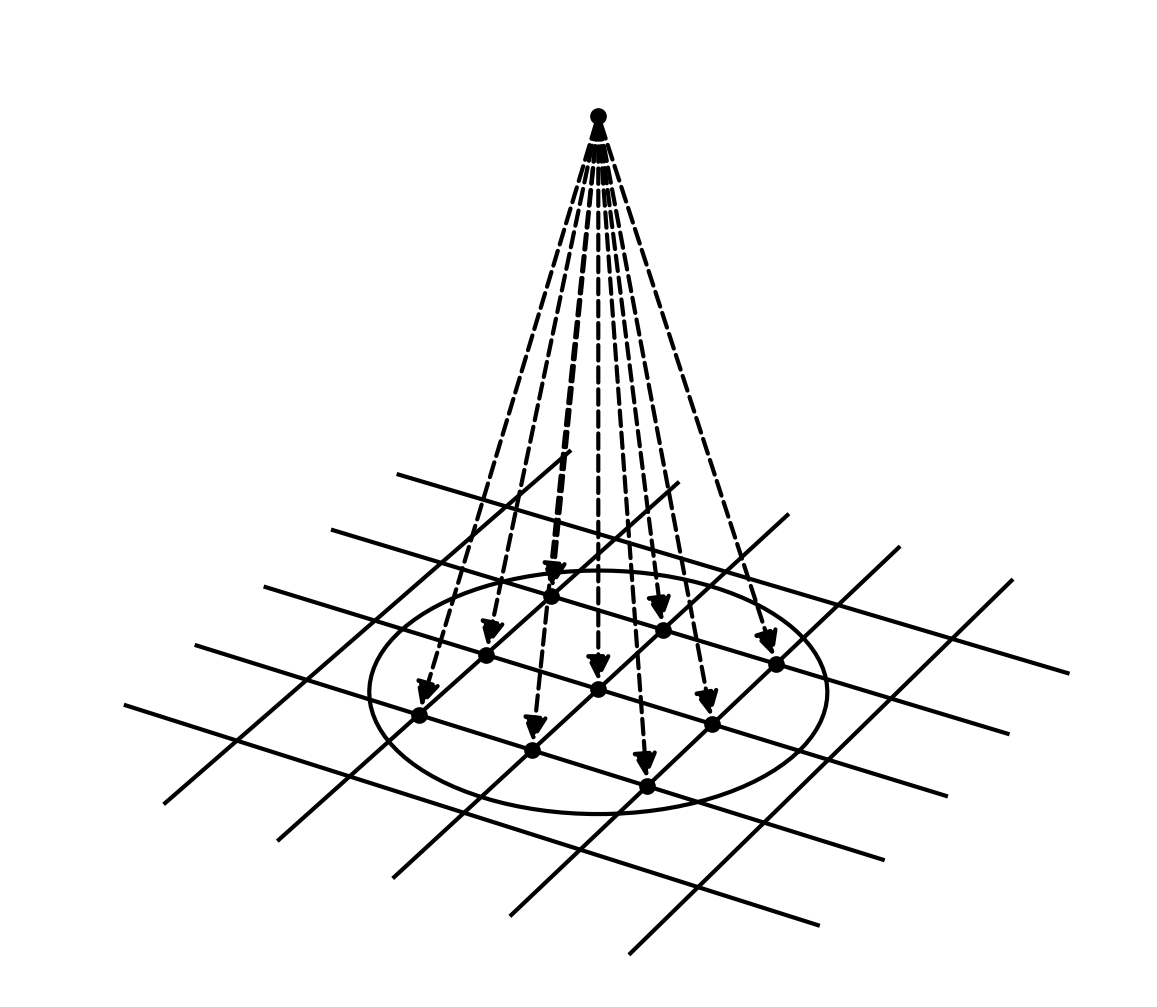}
    \end{minipage}
    \begin{minipage}{0.22\textwidth}
        \centering
        \includegraphics[width=\linewidth]{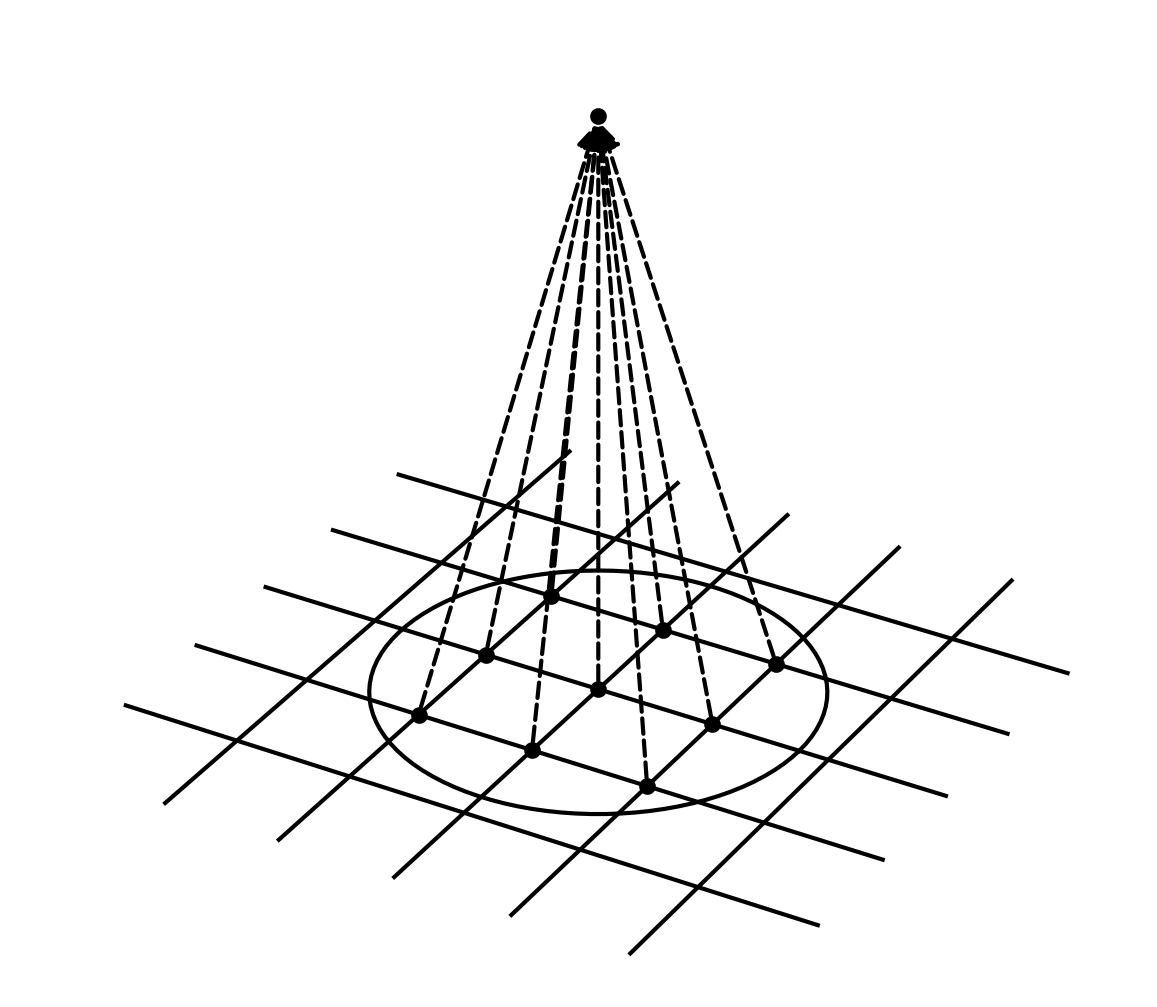}
    \end{minipage}
    \caption[Conceptual illustrations of Projection and Backprojection Cores.]{Conceptual illustrations of Projection and Backprojection Cores. (a) Projection Core: a sample at a non-Cartesian point is projected onto a Cartesian grid. (b) Backprojection Core: samples on a Cartesian grid are backprojected to a non-Cartesian point.}
    \label{fig:gridding-proj_and_backproj}
\end{figure}

1D k-space gridding is expressed as \cite{beatty_rapid_2005}:
\begin{equation}
    m[x] = 
        \mathcal{F}^H\bigg\{ 
                \left( M(k_x) \ast C(k_x) \right) \cdot \mathrm{III}[G k_x]
        \bigg\}  \![x]
        \cdot \left(\frac{1}{c[x]}\right)\label{eq:grid-theory-k_space_gridding_eq}.
\end{equation}
We term the operations within the inverse Fourier transform bracket as the ``Projection Core.'' Within the Projection Core, non-Cartesian k-space data are convolved with an interpolation kernel, $C(k_x)$, and then sampled on an oversampled Cartesian grid. As depicted in Fig.~\ref{fig:gridding-proj_and_backproj}(a), this operation can be visualized as a projection onto a Cartesian grid, with the projection's extent being determined by the width of the kernel.

The implementation of \eqref{eq:grid-theory-k_space_gridding_eq} is outlined as:
\begin{equation}
    M(k_x)
    \rightarrow \boxed{\mathcal{P}^{(k)}}
    \rightarrow \boxed{\mathcal{F}^H}
    \rightarrow \boxed{\mathcal{Z}^H}
    \rightarrow \boxed{\frac{1}{c[x]}}
    \rightarrow m[x].\label{eq:grid-theory-k_space_gridding_process}
\end{equation}
The weighting operation $\boxed{1/c[x]}$ is often referred to as deapodization or preemptive emphasis \cite{bernstein2004handbook, beatty_rapid_2005}. The adjoint of k-space gridding, commonly known as inverse gridding \cite{bernstein2004handbook, beatty_rapid_2005}, is derived by reversing the sequence in \eqref{eq:grid-theory-k_space_gridding_process}:
\begin{equation}
    m[x]
    \rightarrow \boxed{\frac{1}{c^\ast[{x}]}}
    \rightarrow \boxed{\mathcal{Z}}
    \rightarrow \boxed{\mathcal{F}} 
    \rightarrow \boxed{\mathcal{B}^{(k)}}
    \rightarrow M(k_x).\label{eq:grid-theory-k_space_inverse_gridding}
\end{equation}
Here, $\boxed{\mathcal{B}^{(k)}}$ represents the adjoint of the Projection Core. As illustrated in Fig.~\ref{fig:gridding-proj_and_backproj}(b), this operation effectively backprojects values on a Cartesian grid to a non-Cartesian point.

From \eqref{eq:grid-theory-k_space_gridding_process}, we deduce that the k-space gridding operation is essentially a ``corrected'' version of the Projection Core. The convolution in k-space corresponds to multiplication in image-space, adding weight in image-space. The deapodization step neutralizes this weight by dividing by the inverse Fourier transform of $C(k_x)$ \cite{beatty_rapid_2005}. The k-space gridding operator also includes the $\mathcal{Z}^H$ operation. An oversampled grid is generally employed to suppress aliasing artifacts that are produced by the sampling operation in the Projection Core \cite{bernstein2004handbook, beatty_rapid_2005}. However, this oversampled grid in k-space leads to an extended field-of-view in image-space. The $\mathcal{Z}^H$ operation recovers the original image extent by cropping the image center.

k-space gridding offers several advantages. It is accurate, especially when employing a well-designed interpolation kernel \cite{bernstein2004handbook, fessler_nonuniform_2003, fessler_nufft-based_2007, beatty_rapid_2005}. Additionally, optimal kernels and oversampling factors have been extensively studied \cite{fessler_nonuniform_2003,fessler_nufft-based_2007, beatty_rapid_2005}. Moreover, the forward and inverse k-space gridding operators form an exact forward-adjoint pair, making them suitable for iterative reconstructions. Last but not least, both operators can be implemented on GPUs for accelerated computations \cite{gregerson2008implementing}.

\subsection*{Displacement Fields}\label{subsec:grid-theory-field}

Displacement fields are effective for representing nonrigid transformations \cite{rueckert_nonrigid_1999, modersitzki2003numerical}. Consider two images, $m$ and $m_j$, and a nonrigid transform, $\mathcal{T}_j\{\cdot\}$, between them:
\begin{equation}
    m_j = \mathcal{T}_j \{m\}.
\end{equation}
With the displacement field, the nonrigid transform is computed by ``warping'' the input image \cite{rueckert_nonrigid_1999, modersitzki2003numerical}:
\begin{equation}
    m_j
    \begin{bmatrix}
        x\\
        y\\
        z
    \end{bmatrix} = m
    \begin{pmatrix}
        x + d_x[x, y, z]\\
        y + d_y[x, y, z]\\
        z + d_z[x, y, z]
    \end{pmatrix} = m(\vec{\mathrm{r}} + \vec{\mathrm{d}}[\vec{\mathrm{r}}]),\label{eq:grid-theory-dfield}
\end{equation}
where $\vec{\mathrm{d}}$ denotes the displacement field and $\vec{\mathrm{r}} = [x, y, z]^T$ represents a voxel location. Various image registration techniques estimate the displacement field from the two given images \cite{modersitzki2003numerical, vercauteren2007diffeomorphic}.

From \eqref{eq:grid-theory-dfield}, we infer that the voxel locations of the warped image, $m_j$, can be regarded as a set of non-Cartesian points on the reference Cartesian grid of $m$. Therefore, computing $\mathcal{T}_j\{\cdot\}$ is analogous to ``inverse gridding,'' as it synthesizes ``data'' at non-Cartesian points from an image on a Cartesian grid. The key differences are that both the operand and the output are in image-space, and the resulting ``data'' constitutes an image on its own Cartesian grid.

\subsection*{Image-space Gridding}\label{subsec:grid-theory-image_space}

Consider this naive Projection Core for the adjoint of image-space gridding that computes $\mathcal{T}_j^H\{\cdot\}$:
\begin{equation}
    \left( m_j(x) \ast c(x) \right) \cdot \mathrm{III}[N x].
\end{equation}
Here, non-Cartesian data are convolved with an interpolation kernel and subsequently sampled onto a Cartesian grid in image-space. However, this operation introduces two problems. Firstly, convolution in image-space leads to additional weightings in k-space. Secondly, sampling in image-space results in aliasing artifacts in k-space.

Similar to k-space gridding, we define the adjoint of image-space gridding by ``correcting'' these errors:
\begin{equation}
    \hat{m}[x] = \mathcal{F}^{H}\!\!\left\{\!
        \mathcal{F}\!\left\{ 
            % \underbrace{
                \left( m_j(x) \ast c(x)  \right) \!\cdot\! \mathrm{III}[G x]
            % }_{\text{Projection Core}}
        \right\} \!\cdot\! \left(\!\frac{1}{C[k_x]}\!\right)
    \!\right\}\![x],
\end{equation}
where $\hat{m}[x] = \mathcal{T}_j^H\{m_j[x]\}$. Here, we use the deapodization (preemptive emphasis) step and an oversampled Cartesian grid to address the additional weighting and the aliasing, respectively. In other words, the same Projection Core is utilized, but it is computed in image-space with subsequent corrections performed in k-space. The process diagram is outlined as follows:
\begin{equation}
    m_j[x]
    \!\rightarrow\! \setlength{\fboxsep}{2.5pt}\boxed{\mathcal{P}_j^{(i)}}
    \!\rightarrow\! \setlength{\fboxsep}{2.5pt}\boxed{\mathcal{F}}
    \!\rightarrow\! \setlength{\fboxsep}{2.5pt}\boxed{\mathcal{Z}^H}
    \!\rightarrow\! \setlength{\fboxsep}{2.5pt}\boxed{\frac{1}{C[k_x]}}
    \!\rightarrow\! \setlength{\fboxsep}{2.5pt}\boxed{\mathcal{F}^H}
    \!\rightarrow\! \hat{m}[x].\label{eq:grid-theory-im_space_gridding_process-adj}
\end{equation}
With the exception of an additional Fourier transform, this process resembles that of k-space gridding in \eqref{eq:grid-theory-k_space_gridding_process}.

The forward image-space gridding operator that computes $\mathcal{T}\{\cdot\}$ is easily derived by reversing the sequence in \eqref{eq:grid-theory-im_space_gridding_process-adj}:
\begin{equation}
    m[x]
    \!\rightarrow\! \setlength{\fboxsep}{2.5pt}\boxed{\mathcal{F}}
    \!\rightarrow\! \setlength{\fboxsep}{2.5pt}\boxed{\frac{1}{C^\ast[k_x]}}
    \!\rightarrow\! \setlength{\fboxsep}{2.5pt}\boxed{\mathcal{Z}}
    \!\rightarrow\! \setlength{\fboxsep}{2.5pt}\boxed{\mathcal{F}^H}
    \!\rightarrow\! \setlength{\fboxsep}{2.5pt}\boxed{\mathcal{B}_j^{(i)}}
    \!\rightarrow\! m_j[x],\label{eq:grid-theory-im_space_gridding_process-fwd}
\end{equation}
where $m_j[x] = \mathcal{T}_j\{m[x]\}$. With the exception of an additional Fourier transform, this process is analogous to inverse k-space gridding described in \eqref{eq:grid-theory-k_space_inverse_gridding}.

Image-space gridding inherits key benefits from k-space gridding; its two operators form an exact forward-adjoint pair and can be implemented on GPUs for accelerated computations.

Appendix 1 exemplifies the use of image-space gridding in the context of general nonrigid motion correction, where inverting a nonrigid transform is reformulated as a regularized inverse problem.

\section*{Nonrigid SENSE Operators}\label{sec:grid-sense}

\begin{figure}[!htb]
    \centering
    \includegraphics[width=0.47\textwidth, keepaspectratio]{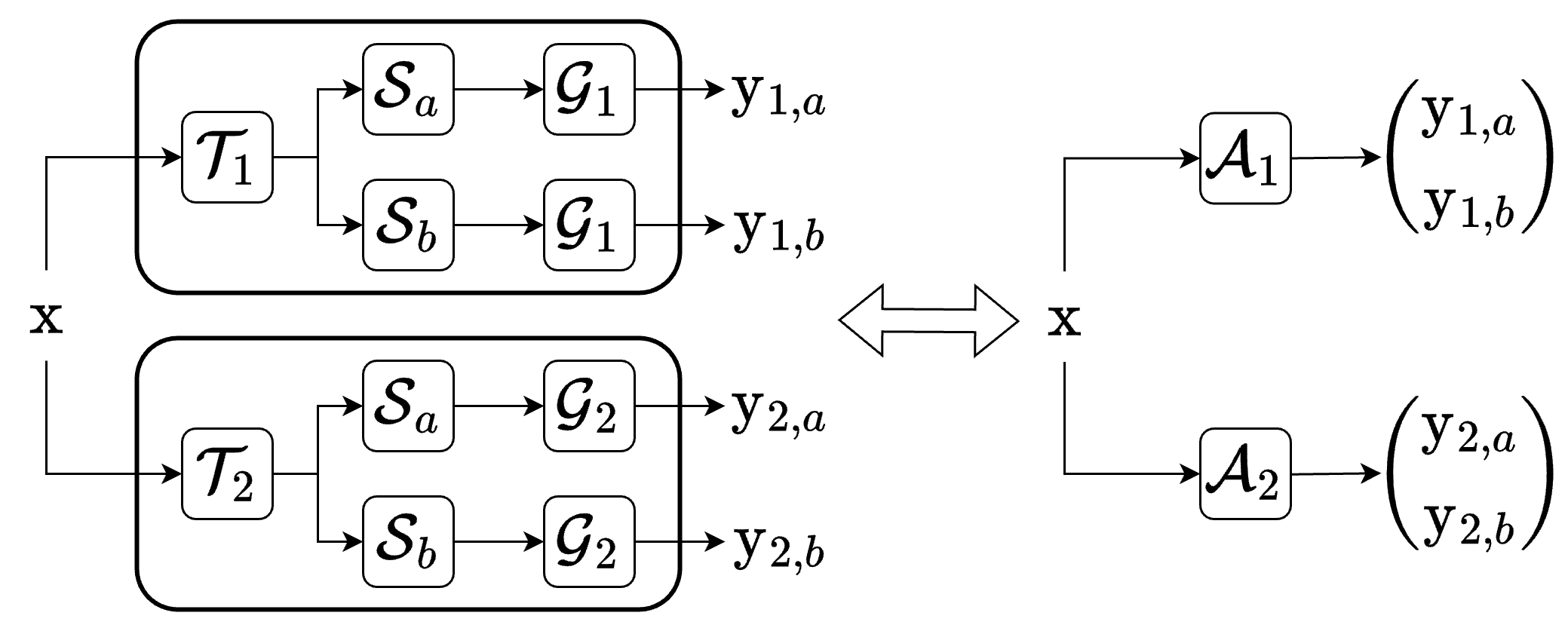}
    \caption[A schematic representation of nonrigid SENSE operator.]{A schematic representation of nonrigid SENSE operators. For simplicity, only two coils, denoted as $a$ and $b$, and two states of the object, labeled as $1$ and $2$, are considered. By vertically stacking both nonrigid SENSE operators and data, we formulate the linear model. $\mathcal{T}$: nonrigid transform, $\mathcal{S}$: coil sensitivity, $\mathcal{G}$: inverse k-space gridding operator, $\mathcal{A}$: nonrigid SENSE operator, $\mathrm{x}$: target object, $\mathrm{y}$: k-space data.}
    \label{fig:grid-sense-stack}
\end{figure}

Here, we propose new operators using image-space gridding, termed nonrigid SENSE operators, to incorporate nonrigid transformations of the object into the multi-coil MR acquisition model. During the prolonged MR scan, we assume that the object exists in several ``states,'' where the object in each state is modeled by the nonrigid transform of the object at a reference state. In cardiac imaging, these states can correspond to different cardiac and/or respiratory phases. For simplicity, consider two coils, denoted as $a$ and $b$, and two states of the object, labeled $1$ and $2$. Let $\mathcal{T}_1\{\cdot\}$ and $\mathcal{T}_2\{\cdot\}$ represent the nonrigid transforms that warp the object $\mathrm{x}$ to State-1 and State-2, respectively. Figure \ref{fig:grid-sense-stack} provides a schematic representation of this example. The nonrigid SENSE operator for State-1, $\mathcal{A}_1\{\cdot\}$, is constructed by preceding the conventional SENSE operator \cite{pruessmann_sense_1999} with the nonrigid transform $\mathcal{T}_1\{\cdot\}$. By vertically stacking both the nonrigid SENSE operators and k-space data, we formulate the data acquisition model, which accounts for the state transitions of the object during the prolonged MR scan, as:
\begin{equation}
    \begin{bmatrix}
        \renewcommand{\arraystretch}{0.8}
        \begin{pmatrix}
            \mathrm{y}_{1, a}\\
            \mathrm{y}_{1, b}
        \end{pmatrix}\\[8pt]
        \renewcommand{\arraystretch}{0.8}
        \begin{pmatrix}
            \mathrm{y}_{2, a}\\
            \mathrm{y}_{2, b}
        \end{pmatrix}
    \end{bmatrix} = \begin{bmatrix}
        \mathrm{y}_{1}\\
        \mathrm{y}_{2}
    \end{bmatrix} = \begin{bmatrix}
        \mathcal{A}_1\\
        \mathcal{A}_2\\
    \end{bmatrix} \mathrm{x},\label{eq:stacked_sense_op}
\end{equation}

\begin{figure}[!htb]
    \centering
    \begin{minipage}{0.47\textwidth}
        \centering
        \includegraphics[width=\linewidth]{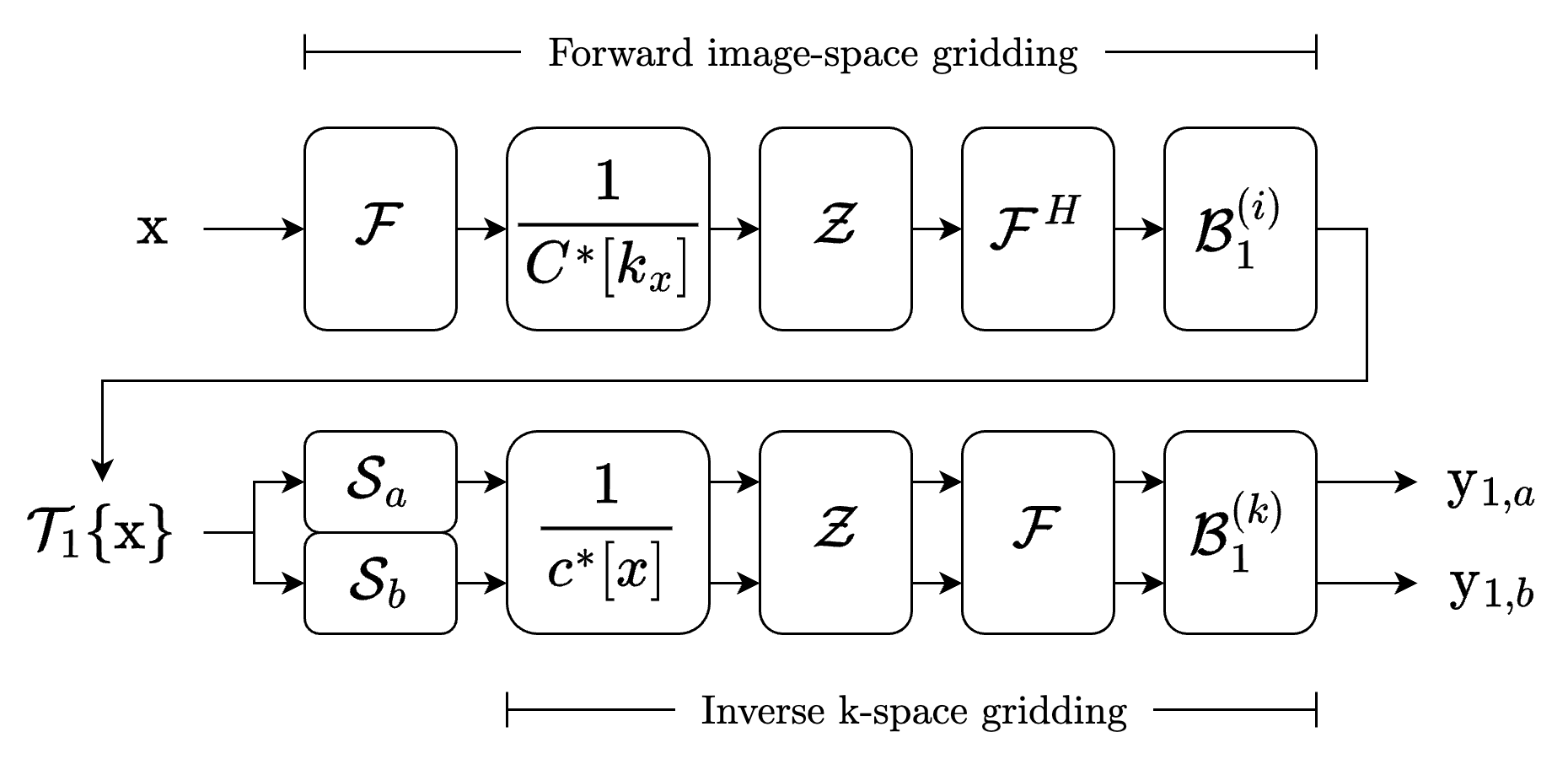}
        \vspace{-18pt}
        \caption*{(a) Forward}
    \end{minipage}\\[5pt]
    \begin{minipage}{0.47\textwidth}
        \centering
        \includegraphics[width=\linewidth]{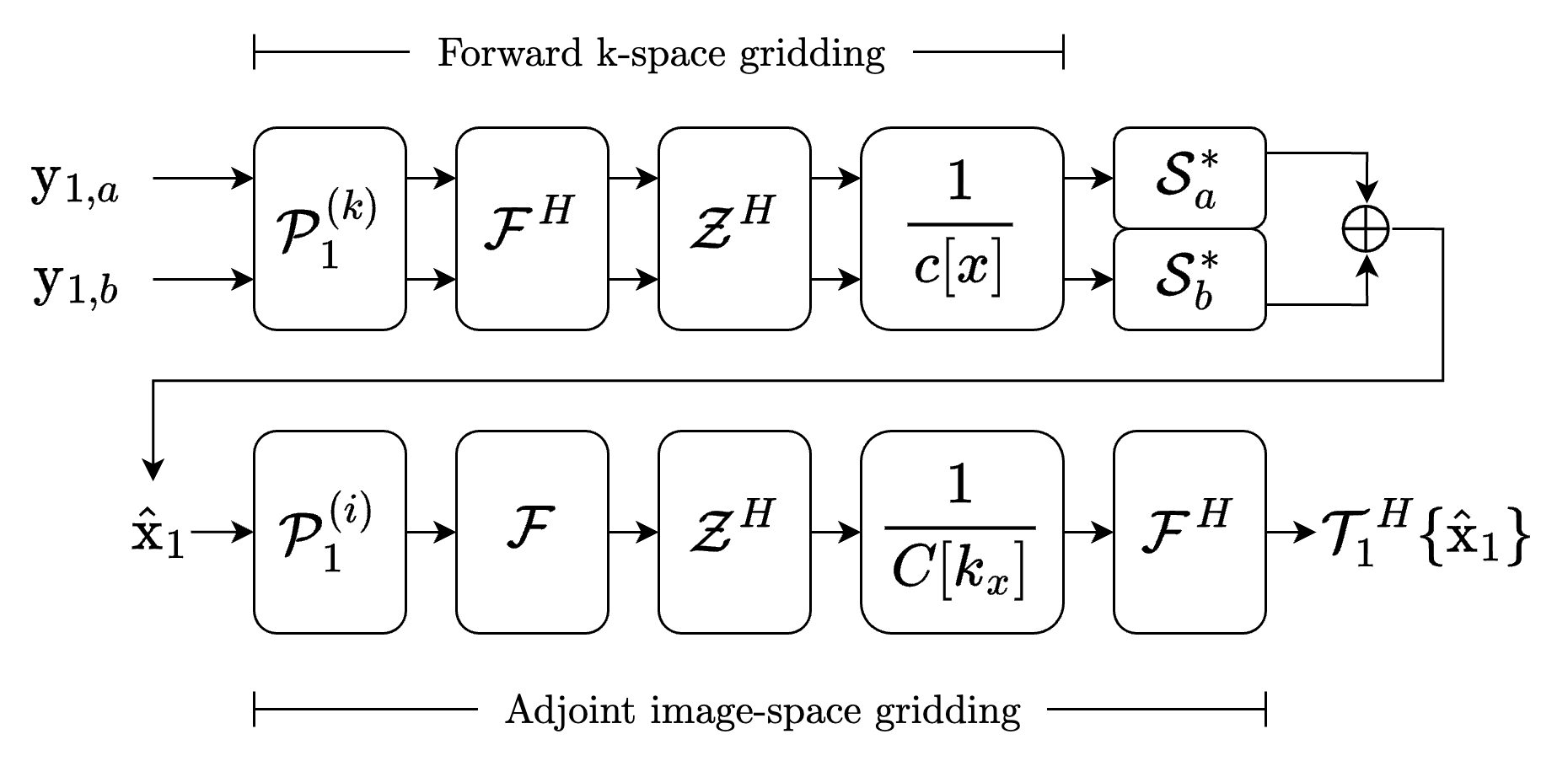}
        \vspace{-18pt}
        \caption*{(b) Adjoint}
    \end{minipage}
    \caption[Process diagrams for forward and adjoint nonrigid SENSE operators.]{Process diagrams for (a) forward and (b) adjoint nonrigid SENSE operators. The nonrigid SENSE operator comprises three serial linear operators: the forward image-space gridding operator, the multiplication by coil sensitivities, and the inverse k-space gridding operator. Its adjoint operator consists of the adjoints of the three linear operators in reversed order: forward k-space gridding operator, the multiplication by the conjugate of coil sensitivities followed by summation, and adjoint image-space gridding operator. $\hat{\mathrm{x}}_1$ denotes the output image from the adjoint of the conventional SENSE operator at State-1.}
    \label{fig:grid-sense_op}
\end{figure}

Given that the nonrigid transform is computed using the forward image-space gridding operator, the nonrigid SENSE operator comprises three sequential linear operators: the forward image-space gridding operator, the multiplication by coil sensitivities, and the inverse k-space gridding operator. Its adjoint operator, on the other hand, consists of the adjoints of these three linear operators in reversed order: the forward k-space gridding operator, the multiplication by the conjugate of coil sensitivities followed by summation, and the adjoint image-space gridding operator. The process diagrams for the forward and adjoint nonrigid SENSE operators are depicted in Fig.~\ref{fig:grid-sense_op}. %Note that two sets of non-Cartesian locations are employed to construct $\mathcal{A}_a\{\cdot\}$: the voxel locations warped by the displacement field, used for the forward image-space gridding operator, and a subset of the sampling trajectory, used for the inverse k-space gridding operator.

\section*{Motion Capturing for Segmented Data Acquisitions}

In free-breathing cardiac MR imaging, cardiac triggering is commonly used to synchronize the pulse sequence with the subject's cardiac cycle \cite{bernstein2004handbook}. A subset of data is acquired during each heartbeat under the assumption that cardiac motion remains static during a specific cardiac phase determined by a trigger delay \cite{bernstein2004handbook, hofman1998quantification, wang2001coronary}. In our motion capturing scheme for the segmented data acquisition, data are acquired along two types of non-Cartesian trajectories during each heartbeat: a subset of a high-resolution trajectory that sparsely covers 3D k-space, followed by a full low-resolution 3D trajectory.

We reconstruct 3D iNAVs \cite{wu2013free, addy20173d} using data near the origin of k-space to estimate bulk motion. The process of 3D translational motion correction is summarized as follows: First, one of the 3D iNAVs is selected as a reference. Translational motion for each heartbeat is then estimated by computing correlations between this reference and the other 3D iNAVs \cite{beare_image_2018, mccormick2014itk, lowekamp2013design}. Finally, the estimated bulk translational motion is corrected by applying linear phase shifts to the k-space data.

At various respiratory phases, the heart may undergo different nonrigid transformations. In other words, data are collected at several distinct states of the imaging object (referred to as subscripts $1$ and $2$ in \eqref{eq:stacked_sense_op}), where these states correspond to respiratory phases in free-breathing cardiac MR imaging. Based on the estimated bulk motion, we identify the respiratory phase of each heartbeat, assuming the location of the heart is closely related to the respiratory phase. Since the high-resolution data from each heartbeat are distributed across 3D k-space, binning these data into discrete respiratory phases enables the reconstruction of high-resolution 3D self-iNAVs \cite{pang2014whole} for individual respiratory phases. In this way, we capture nonrigid motion using a relatively short low-resolution iNAV acquisition during each heartbeat.

We choose one of these 3D self-iNAVs as a reference and employ an established nonrigid image registration method to calculate displacement fields relative to this reference. Using the estimated displacement fields, we incorporate the nonrigid transform due to respiratory motion into the MR acquisition model through the nonrigid SENSE operator. By stacking pairs of k-space data and their corresponding nonrigid SENSE operators, we formulate the following regularized inverse problem for nonrigid motion-corrected reconstruction:
\begin{equation}
    \operatorname*{arg\,min}_{\mathrm{x}} 
    \left\lVert\begin{bmatrix}
        \mathrm{y}_0\\
        \vdots\\
        \mathrm{y}_{N - 1}
    \end{bmatrix} - \begin{bmatrix}
        \mathcal{A}_0 \\
        \vdots\\
        \mathcal{A}_{N - 1}
    \end{bmatrix}\mathrm{x}\right\rVert_2^2 + \lambda \|\mathcal{W} \mathrm{x}\|_1,\label{eq:recon-resp-optimization}
\end{equation}
where $N$ is the number of respiratory phases, $\mathrm{y}_i$ and $\mathcal{A}_i$ denote the multi-coil data and the nonrigid SENSE operator at the $i$-th respiratory phase, respectively, and $\mathcal{W}$ is a sparsifying operator, such as a wavelet transform \cite{lustig2007sparse}.

% The low-resolution 3D iNAVs are used to estimate translational motion. Subsequently, we bin data from several heartbeats to reconstruct high-resolution 3D self-navigating image-based navigators (3D self-iNAVs) \cite{pang2014whole} for estimating nonrigid motion across different respiratory phases. This scheme minimizes the acquisition footprint for motion capturing while effectively capturing complex nonrigid motion.

\section*{Nonrigid Respiratory Motion Correction}

\begin{figure}[!htb]
    \centering
    \includegraphics[width=0.47\textwidth, keepaspectratio]{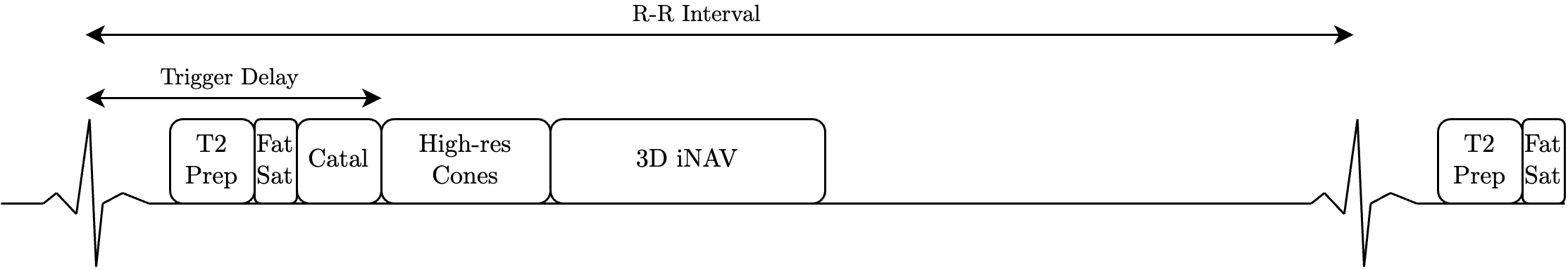}
    \caption[Imaging modules over an R-R interval for single-phase imaging.]{Imaging modules over an R-R interval for single-phase imaging. T2 Prep: T2 preparation, Fat Sat: fat saturation, Catal: catalyzation, High-res Cones: main data acquisition window, 3D iNAV: 3D image-based navigator.}
    \label{fig:recon-resp-imaging_modules}
\end{figure}

For free-breathing coronary MR angiography, we employ a data acquisition scheme illustrated in Fig.~\ref{fig:recon-resp-imaging_modules} using a balanced steady-state free precession (bSSFP) sequence, similar to prior studies \cite{addy_high-resolution_2015, malave_whole-heart_2019, wu2013free}. During each R-R interval, high-resolution data are collected along 18 conic interleaves, which are a subset of a 3D Cones phyllotaxis trajectory designed for 1.2-mm isotropic resolution \cite{malave_whole-heart_2019}. This primary imaging module is followed by a 3D iNAV acquisition using a 3D Cones trajectory designed for 4.4-mm isotropic resolution \cite{addy_high-resolution_2015}. The following contrast-preparation modules are applied before data acquisition: T2 preparation \cite{brittain_coronary_1995}, fat saturation, and sinusoidal bSSFP catalyzation.

\begin{figure}[!htb]
    \centering
    \includegraphics[width=0.47\textwidth, keepaspectratio]{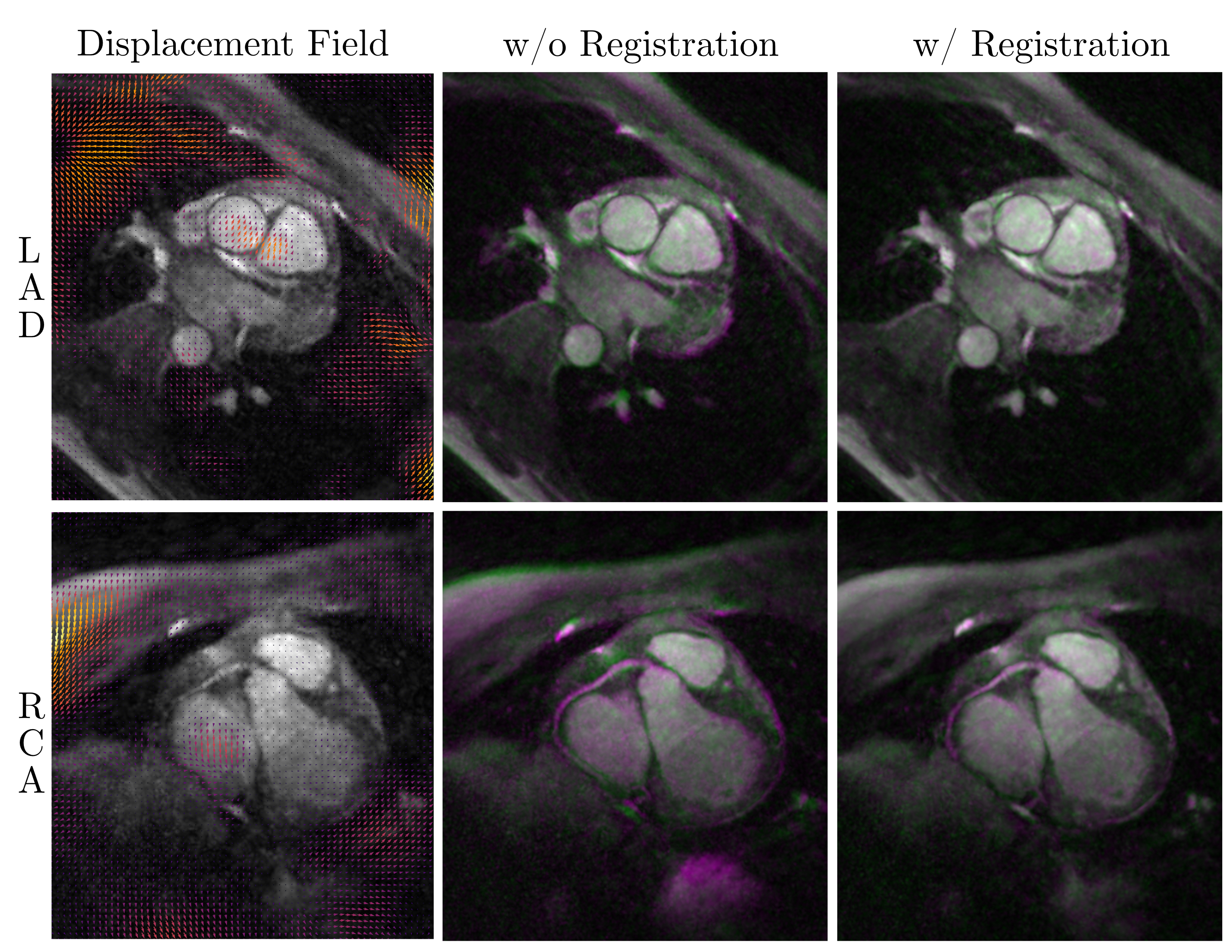}
    \caption[Estimated nonrigid motion.]{Left: Estimated displacement fields and 3D self-iNAVs. Middle and right: The reference and one of the non-reference 3D self-iNAVs overlaid in purple and green, respectively, before (middle) and after (right) registration. Gray pixels signify well-aligned areas, while colored pixels highlight discrepancies. This comparison underscores the heart's nonrigid transformations across different respiratory phases, illustrating the advantages of nonrigid motion-corrected reconstructions.}
    \label{fig:recon-resp-registration-1}
\end{figure}

\begin{figure}[!htb]
    \centering
    \includegraphics[width=0.47\textwidth, keepaspectratio]{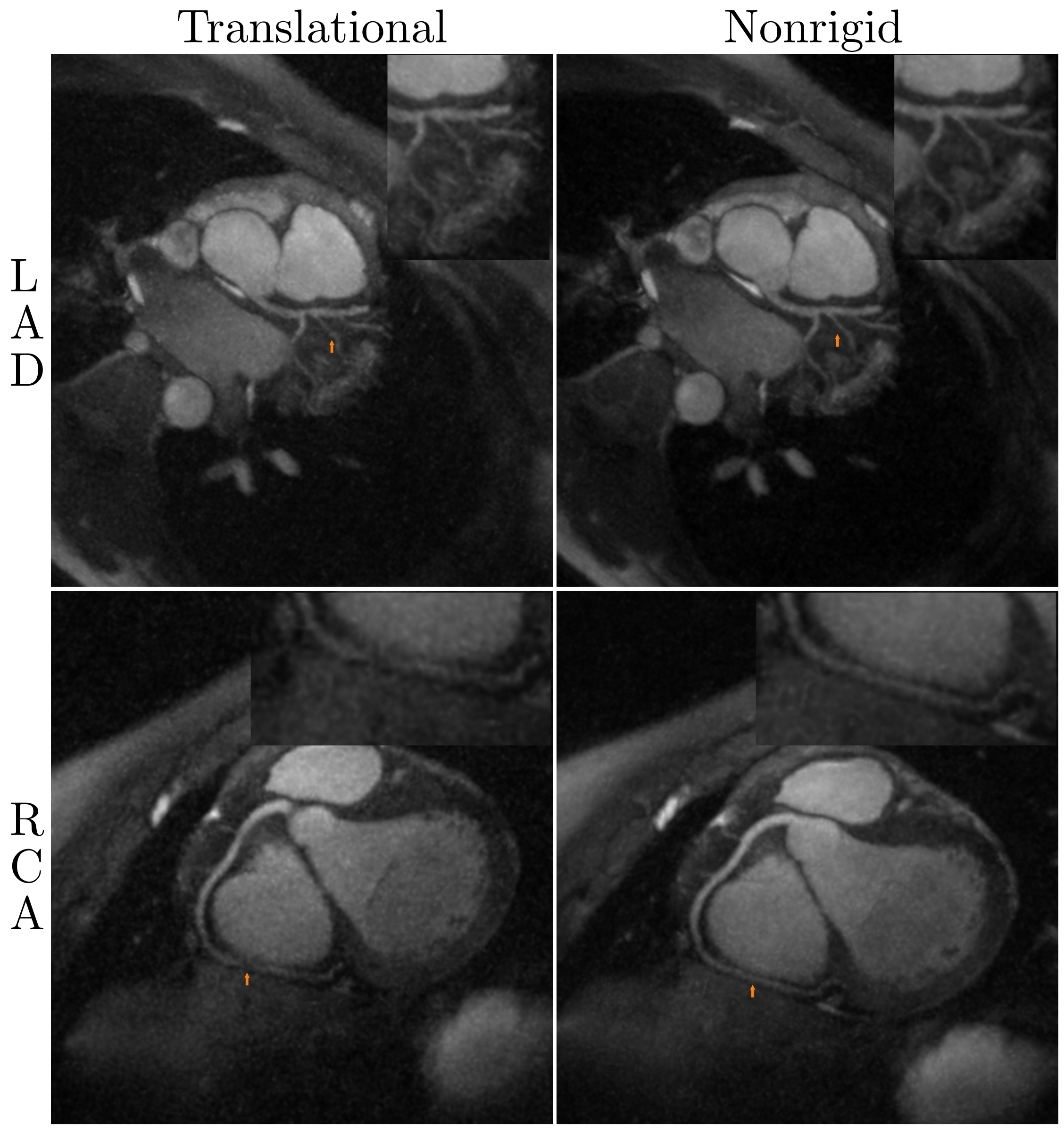}
    \caption[Reconstructed slices near the LAD and RCA (Subject 1).]{Reconstructed slices near the LAD and RCA (Subject 1). The proposed method exhibited improved sharpness not only near the coronary arteries (orange arrows) but also in non-cardiac regions compared to 3D translational motion-corrected reconstruction.}
    \label{fig:recon-resp-recon-1}
\end{figure}

\begin{figure}[!htb]
    \centering
    \includegraphics[width=0.47\textwidth, keepaspectratio]{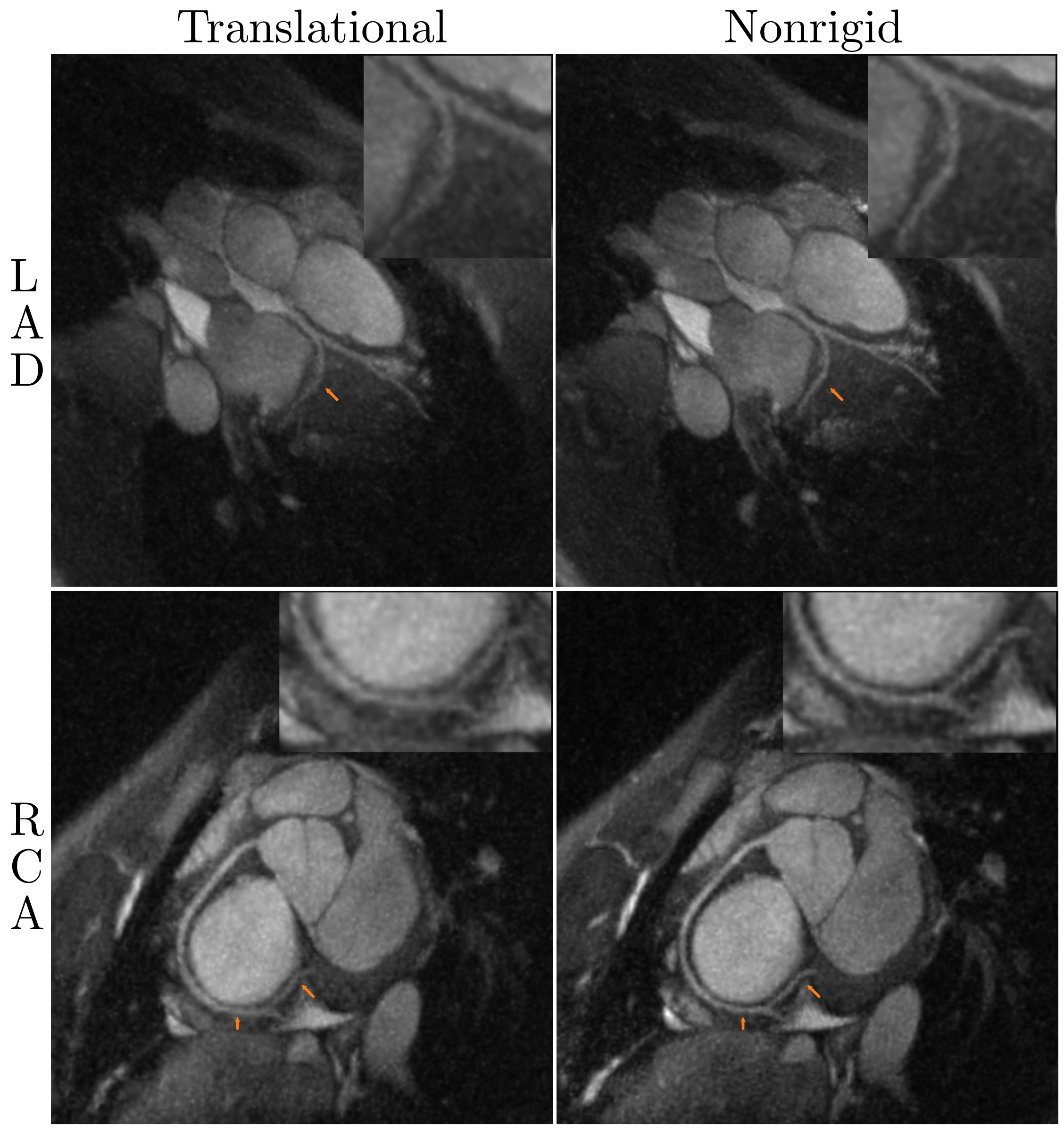}
    \caption[Reconstructed slices near the LAD and RCA (Subject 2).]{Reconstructed slices near the LAD and RCA (Subject 2). The proposed method exhibited improved sharpness not only near the coronary arteries (orange arrows) but also in non-cardiac regions compared to 3D translational motion-corrected reconstruction.}
    \label{fig:recon-resp-recon-2}
\end{figure}

Figures~\ref{fig:recon-resp-registration-1}--\ref{fig:recon-resp-recon-2} present the preliminary results. Data were acquired on a 1.5T GE Signa scanner using an 8-channel cardiac coil \cite{malave_whole-heart_2019} from two healthy subjects.

We identified four respiratory phases using k-means clustering based on the translational motion estimates. The end-expiration phase was selected as the reference for estimating nonrigid motion, as the corresponding cluster showed the least variance in translational motion \cite{jang2021nonrigid}. For nonrigid motion estimation, we employed the symmetric forces diffeomorphic demons algorithm \cite{vercauteren2007diffeomorphic}, implemented in \verb|SimpleITK| \cite{beare_image_2018, mccormick2014itk, lowekamp2013design}. To solve \eqref{eq:recon-resp-optimization}, we used the Wavelet-FISTA algorithm \cite{beck2009fast}.

To validate the measured nonrigid motion, we overlaid the reference and one of the non-reference 3D self-iNAVs in purple and green, respectively, as shown in the middle and rightmost columns of Fig.~\ref{fig:recon-resp-registration-1} \cite{jang2021nonrigid}. With this color scheme, well-aligned areas appear gray due to the even mixture of red, blue, and green, while colored pixels highlight discrepancies between the reference and the non-reference respiratory phase. Figure~\ref{fig:recon-resp-registration-1} illustrates the nonrigid transformations of the heart across different respiratory phases, demonstrating the potential benefits of nonrigid motion-corrected reconstruction.

Figures~\ref{fig:recon-resp-recon-1} and \ref{fig:recon-resp-recon-2} show reconstructed slices near the left anterior descending artery (LAD) and the right coronary artery (RCA) \cite{jang2021nonrigid}. These results demonstrate that the proposed method not only enhances sharpness near the coronary arteries but also improves image quality in non-cardiac regions, outperforming translational motion-corrected reconstruction \cite{jang2021nonrigid}.

% \begin{SCfigure*}[\sidecaptionrelwidth][t!]
%     \centering
%     \includegraphics[width=11.0cm, keepaspectratio]{figures/Respiration/data_1-recon.png}
%     \caption[Reconstructed slices near LAD and RCA (Subject 1).]{Reconstructed slices near LAD and RCA (Subject 1). The proposed method exhibited improved sharpness not only near the coronary arteries (orange arrows) but also in non-cardiac regions compared to 3D translational motion-corrected reconstruction.}
%     \label{fig:recon-resp-recon-1}
% \end{SCfigure*}

% \begin{SCfigure*}[\sidecaptionrelwidth][t!]
%     \centering
%     \includegraphics[width=11.0cm, keepaspectratio]{figures/Respiration/data_0-recon.png}
%     \caption[Reconstructed slices near LAD and RCA (Subject 2).]{Reconstructed slices near LAD and RCA (Subject 2). The proposed method exhibited improved sharpness not only near the coronary arteries (orange arrows) but also in non-cardiac regions compared to 3D translational motion-corrected reconstruction.}
%     \label{fig:recon-resp-recon-2}
% \end{SCfigure*}

% \clearpage % For draft

\section*{Conclusion}

We derived a new pair of linear operators, called image-space gridding operators, to compute the forward and adjoint of a nonrigid transform. We then introduced the nonrigid SENSE operator to integrate the nonrigid transform into the MR acquisition model. By stacking multiple nonrigid SENSE operators for different object states, we formulated an MR acquisition model that accounts for the object's nonrigid transformations during the prolonged MR scan.

We applied the proposed approach to free-breathing coronary MR angiography. During each heartbeat, two sets of data were acquired: segmented high-resolution data distributed across 3D k-space and low-resolution data. For each heartbeat, the low-resolution 3D iNAVs were reconstructed to estimate translational motion and identify the respiratory phase. For each respiratory phase, the high-resolution 3D self-iNAVs were reconstructed using the binned high-resolution data. Nonrigid respiratory motion was estimated by comparing the 3D self-iNAVs and was subsequently incorporated into the MR acquisition model through the nonrigid SENSE operators. In short, using image-space gridding and the proposed motion estimation process, we reformulated the nonrigid respiratory motion-corrected reconstruction as a standard regularized inverse problem.

In a preliminary study, we demonstrated that the proposed method enhances voxel depiction and improves image quality in non-cardiac regions, outperforming translational motion-corrected reconstruction. Future work will involve a more comprehensive comparison using additional in-vivo datasets.

\section*{Acknowledgments}

This study was supported by NIH Grant R01 HL127039. Portions of this work were derived from K.E.J.'s doctoral thesis submitted to Stanford University, available at \href{https://purl.stanford.edu/sc219dv4593}{https://purl.stanford.edu/sc219dv4593}.

% \acknow{This study was funded by NIH Grant R01 HL127039. Portions of this work were derived from K.E.J.'s doctoral thesis submitted to Stanford University, which is available at \href{https://purl.stanford.edu/sc219dv4593}{https://purl.stanford.edu/sc219dv4593}.}
% \showacknow

\section*{Appendix 1: Inverting a Nonrigid Transform}\label{sec:grid-toy}

\begin{figure}[!htb]
    \centering
    \begin{minipage}{0.23\textwidth}
        \centering
        \includegraphics[width=\linewidth]{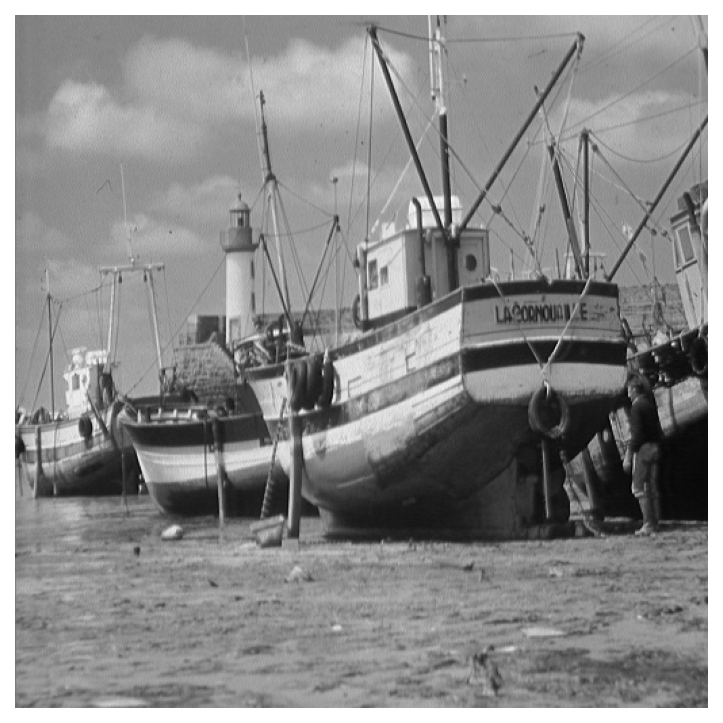}
        \caption*{(a) Ground-truth image}
    \end{minipage}
    \begin{minipage}{0.23\textwidth}
        \centering
        \includegraphics[width=\linewidth]{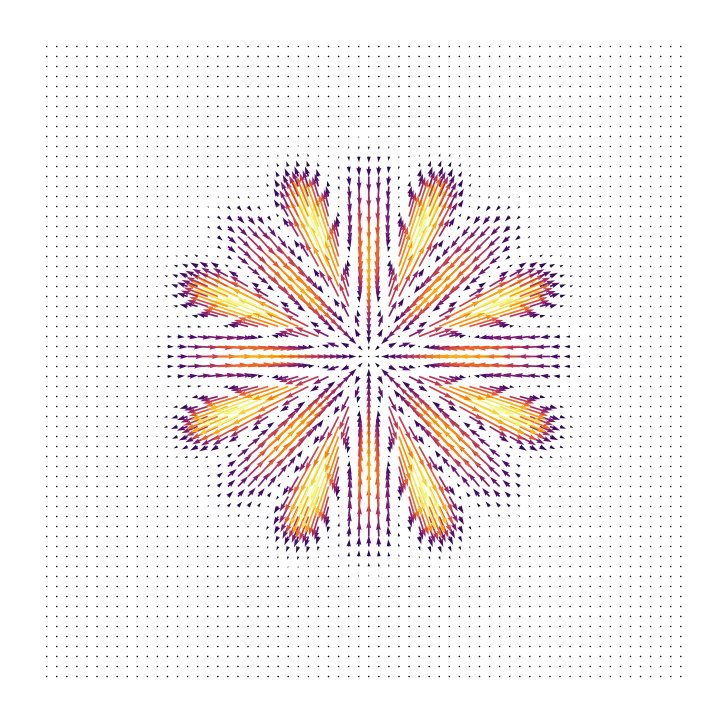}
        \caption*{(b) Displacement field}
    \end{minipage}
    \caption[Ground-truth image and displacement field.]{(a) Ground-truth image and (b) displacement field for the numerical study.}
    \label{fig:grid-toy-fixed}
\end{figure}

In this Appendix, we consider inverting a nonrigid transform in image-space, exemplifying the use of image-space gridding in the context of general nonrigid motion correction. We have a ground-truth image $\mathrm{p}[x, y]$ and a displacement field $\vec{\mathrm{d}}[x, y]$, as illustrated in Fig.~\ref{fig:grid-toy-fixed}. Let $\mathcal{T}\{\cdot\}$ denote a nonrigid transform defined by the displacement field. Let $\mathrm{q}[x, y]$ denote the image after this nonrigid transform:
\begin{equation}
    \mathrm{q}[x, y] = \mathcal{T} \{ \mathrm{p}[x, y] \}.
\end{equation}

\begin{figure}[!htb]
    \centering
    \begin{minipage}{0.23\textwidth}
        \centering
        \includegraphics[width=\linewidth]{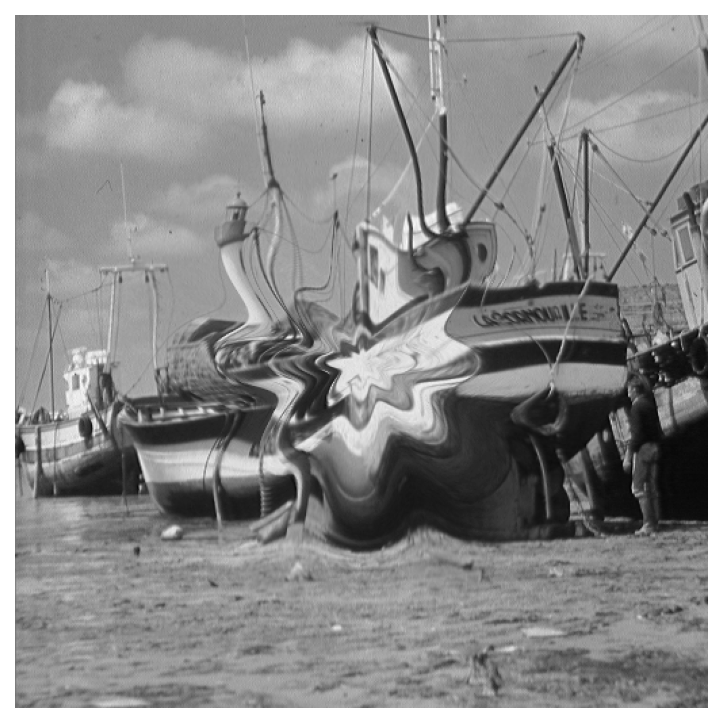}
        \caption*{(a) SimpleITK}
    \end{minipage}
    \begin{minipage}{0.23\textwidth}
        \centering
        \includegraphics[width=\linewidth]{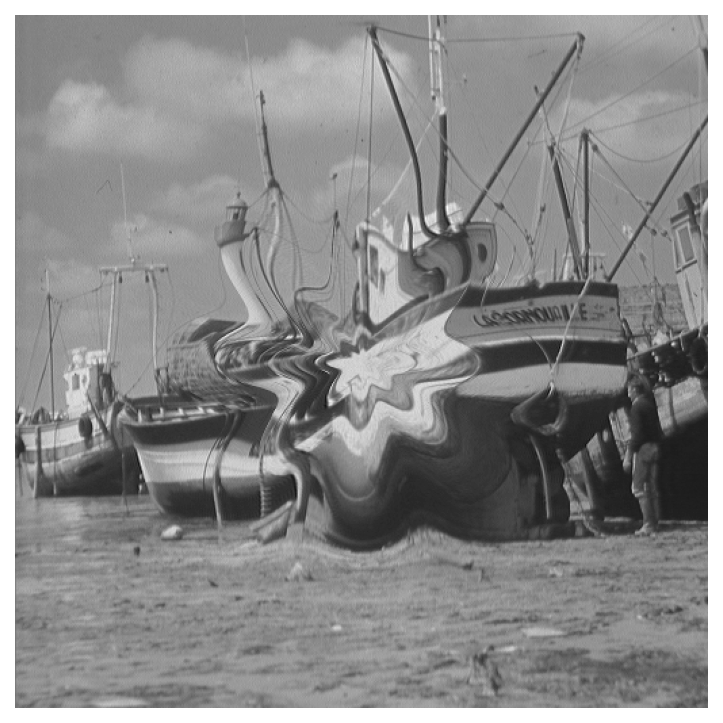}
        \caption*{(b) Image-space gridding}
    \end{minipage}
    \caption[Warped images using SimpleITK and image-space gridding]{Transformed images computed using (a) SimpleITK and (b) image-space gridding. The results of both approaches are visually identical.}
    \label{fig:grid-toy-moving}
\end{figure}

The first task is to synthesize $\mathrm{q}[x, y]$. We can use \verb|SimpleITK| \cite{beare_image_2018} to apply the nonrigid transform to the ground-truth image $\mathrm{p}[x, y]$. Alternatively, we can employ the forward image-space gridding operator to compute $\mathrm{q}[x, y]$. The results of both approaches are visually identical, as shown in Fig.~\ref{fig:grid-toy-moving}.

\begin{figure}[!htb]
    \centering
    \begin{minipage}{0.23\textwidth}
        \centering
        \includegraphics[width=\linewidth]{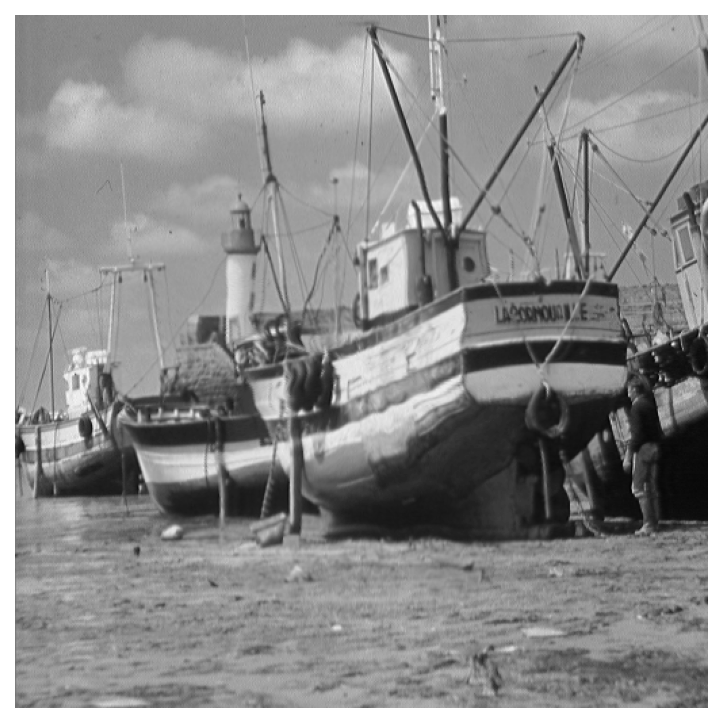}
        \caption*{(a) Inverted displacement field}
    \end{minipage}
    \begin{minipage}{0.23\textwidth}
        \centering
        \includegraphics[width=\linewidth]{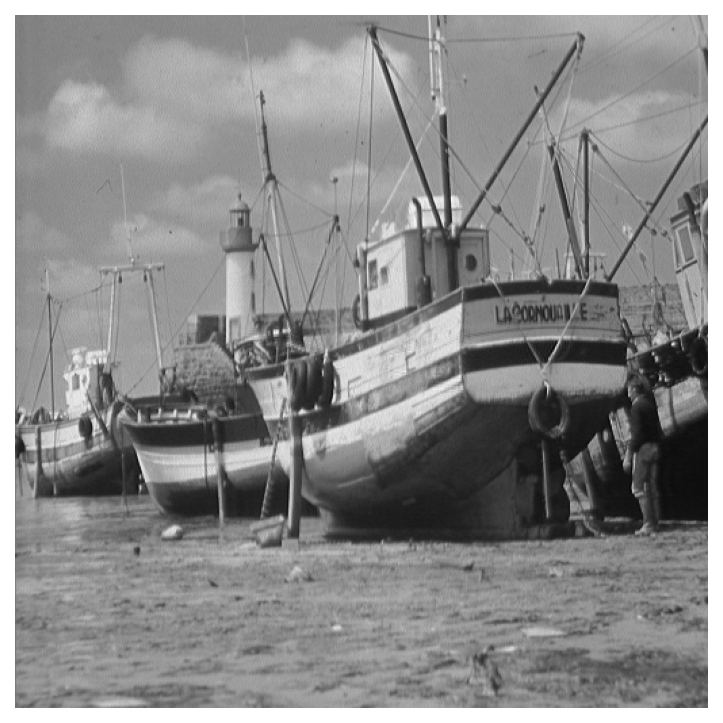}
        \caption*{(b) Iterative approach}
    \end{minipage}
    \caption[Recovered images using the naive and the iterative approaches.]{Recovered images using (a) an inverted displacement field, $(-\vec{\mathrm{d}})$, and (b) wavelet-FISTA with image-space gridding, illustrating the challenges in inverting a nonrigid transform. The iterative approach showed a much more accurate recovery.}
    \label{fig:grid-toy-recon}
\end{figure}

The second task is to recover the original image from $\mathrm{q}[x, y]$ assuming that the ground-truth displacement field, $\vec{\mathrm{d}}[x, y]$, is available. A naive approach is to negate the image warping by applying the inverted displacement field, $-\vec{\mathrm{d}}[x, y]$, to $\mathrm{q}[x, y]$. Figure~\ref{fig:grid-toy-recon}(a) shows that inverting a nonrigid transform is not straightforward \cite{chen2008simple} even if the exact displacement field is given.

Alternatively, we recover $\mathrm{p}[x, y]$ by solving this optimization problem:
\begin{equation}
    \operatorname*{arg\,min}_{\mathrm{p}} \| \mathrm{q} - \mathcal{T}\{ \mathrm{p} \} \|_2^2 + \lambda \|\mathcal{W}\{\mathrm{p}\}\|_1,\label{eq:grid-toy-problem}
\end{equation}
where $\mathcal{W}\{\cdot\}$ denotes a sparsifying transform \cite{lustig2007sparse} such as a wavelet transform. Using image-space gridding, we construct an exact forward-adjoint pair of linear operators to represent the nonrigid transform and its adjoint, $\mathcal{T}\{\cdot\}$ and $\mathcal{T}^H\{\cdot\}$. Subsequently, we solve \eqref{eq:grid-toy-problem} using an iterative solver. For this example, we used 400 iterations of wavelet-FISTA \cite{beck2009fast} with $\lambda=10^{-6}$. As shown in Fig.~\ref{fig:grid-toy-recon}(b), this approach provides a more accurate recovery.

\bibsplit[15]
%Use \bibsplit to split the references from the body of the text. Value "[2]" represents the number of reference in the left column (Note: Please avoid single column figures & tables on this page.)

% Bibliography
% \bibliography{reference}

\end{document}